\documentclass[aps,prl,twocolumn]{revtex4}
\usepackage{graphicx}
\usepackage{amsmath}
\usepackage{epsfig}
\begin{document}
\title{Exact results for spin dynamics and fractionization in the Kitaev Model}
\author{G. Baskaran, Saptarshi Mandal and R. Shankar}
  \affiliation{The Institute of Mathematical Sciences, CIT Campus,
  Chennai 600 113, India.}

\begin{abstract}

We present certain exact analytical results for dynamical spin correlation 
functions in the Kitaev Model. It is the first result of its kind in 
non-trivial quantum spin models. The result is also novel: in spite of presence 
of gapless propagating Majorana fermion excitations, dynamical two spin 
correlation functions are identically zero beyond nearest neighbor separation. 
This shows existence of a gapless but short range spin liquid. An unusual, 
\emph{all energy scale fractionization } of a spin -flip quanta, into two 
infinitely massive $\pi$-fluxes and a dynamical Majorana fermion, is shown to 
occur. As the Kitaev Model exemplifies topological quantum computation, our 
result presents new insights into qubit dynamics and generation of topological 
excitations.
\end{abstract}

\maketitle
In the field of quantum computers and quantum communications, practical 
realizations of qubits that are robust and escape decoherence is a foremost 
challenge\cite{general}. In this context Kitaev proposed\cite{kitaev1} certain 
emergent topological excitations in strongly correlated quantum many body 
systems as robust qubits. In a fault tolerant quantum computation 
scheme\cite{kitaev1,preskill,friedman}, Kitaev constructed a non-trivial and 
exactly solvable 2-dimensional spin model\cite{kitaev1} and illustrated basic 
ideas. In some limit it also becomes the celebrated `toric code' Hamiltonian. 
The Kitaev model has come closer to reality, after recent proposals for 
experimental realizations\cite{demler,zoller}  and schemes for manipulation 
and detection\cite{dasSarmaOptLatt}. In initialisation, error 
correction and read out operations, it is `spins' rather than 
emergent topological degrees of freedom that are directly accessed from outside.
Thus an understanding of dynamic spin correlations is of paramount 
importance. 

We present certain exact analytical results for time dependent spin 
correlation functions, in arbitrary eigen-states of the Kitaev Model. Our 
results are non-trivial and novel, with possible implications for new 
quantum computational schemes. Further our result is unique in the sense that 
it is the first exact result for equilibrium dynamical spin correlation 
functions in a non trivial 2D quantum spin model. 

We show that dynamical two spin correlation functions are short ranged and 
vanish identically beyond nearest neighbor sites for all time t, for all 
values of the coupling constants $J_x, J_y$ and $J_z$, even in the domain 
of J's where the model is gapless. Our result shows rigorously that it is a 
short range quantum spin liquid and long range spin order is absent. 
We obtain a compact form for the time dependence, which makes the physics 
transparent.

Kitaev Model is known to support dynamical Majorana fermions and 
static $\pi$-flux eigen-excitations. We show how fractionization 
\cite{frac,bza} of a local spin--flip quanta into a bound
pair of static $\pi$-flux excitations and a free Majorana fermion 
occurs.

In the present paper we have restricted our calculation to dynamical 
correlation functions for time independent Hamiltonians, in arbitrary 
eigen-states and thermal states. In actual quantum computations, key 
manipulations such as braiding involve parametric change of the Hamiltonian 
and adiabatic transport of topological degrees of freedom\cite{dasSarmaOptLatt}.
In principle, some of the needed `non equilibrium' dynamical correlation 
functions may be obtained by convolution of our results with suitable Berry 
phase factors. 

In our work we follow Kitaev\cite{kitaev1} and use the Majorana fermion 
representation of spin-half operators and an enlarged Hilbert space. What is 
remarkable is that, because of the presence of certain local conserved 
quantities in the Kitaev Model, Hilbert space enlargement only produces 
`gauge copies', without altering the energy spectrum. This luxury is absent 
for standard 2D Heisenberg models when studied using enlarged fermionic Hilbert
space \cite{bza,bss}. 

The Kitaev Hamiltonian is
\begin{equation}
\label{ham1}
H=-J_x\sum_{\langle ij\rangle_x}~\sigma^x_i\sigma^x_j
-J_y\sum_{\langle ij\rangle_y}~\sigma^y_i\sigma^y_j
-J_z\sum_{\langle ij\rangle_z}~\sigma^z_i\sigma^z_j
\end{equation}
where $i,j$ label the sites of a hexagonal lattice, 
$\langle ij\rangle_a,~a=x,y,z$ 
denotes the nearest neighbor bonds in the $a$'th direction. 
The model has no continuous global spin symmetry. All bond interactions are 
Ising like, albeit in different quantisation directions x, y and z, in three 
different bond types, making the model quantum mechanical. Further, it renders 
a high degree of frustration; that is, even at a classical level a given spin 
can not satisfy conflicting demands, from 3 neighbors, of orientations in 
mutually orthogonal directions.
\begin{figure}[h!]
\center{\hbox{\epsfig{figure=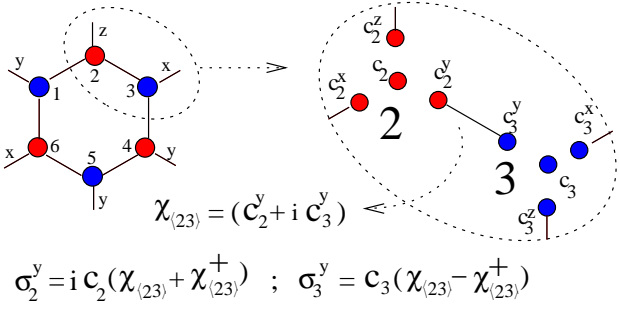}}}
\caption{Elementary hexagon and `bond fermion' construction. A spin is replaced with 4 majorana
fermions ($c,c^{x},c^{y},c^{z}$).  Bond fermion $\chi_{\langle23\rangle}$ for the bond joining site 2 and site 3 is shown . Spin operators are also defined.}
\end{figure}
The model has a rich local symmetry. A specific product of 6 spin components in every elementary hexagon, $\sigma^y_1\sigma^z_2\sigma^x_3\sigma^y_4\sigma^z_5\sigma^x_6$ (figure 1) commutes with the full Hamiltonian. Thus there is one conserved $Z_2$ charge $\pm 1$, at every dual lattice site of the hexagonal lattice. The model is exactly solvable and becomes non interacting Majorana fermions, propagating in the background of static $Z_2$ gauge fields. Different possible $Z_2$ charges separate the Hilbert space into super selected sectors. The ground state corresponds to all $Z_2$ charges = + 1.  In this sector, for a range of J's, Majorana fermions are gapless, including the special point
$J_x = J_y = J_z$.

Following Kitaev, we represent the spins in terms of Majorana fermions. At each
site, we define 4 Majorana fermions, $c^\alpha,~\alpha=0,x,y,z$,
\begin{equation}
\label{mfacr}
\{c^\alpha,c^\beta\}=2\delta_{\alpha\beta}
\end{equation}
Four Majorana (real) fermions make two complex fermions, making the Hilbert 
space 4 dimensional. Notionally, Hilbert space dimension of a Majorana
fermion is $\sqrt 2$, an irrational number, reminding us that Majorana 
fermions have to occur in pairs (leading to a $\sqrt2\times\sqrt2 = 2$
dimensional Fock space) in physical problems.

The dimension of Hilbert space of N spins is $2^N$.
The enlarged Hilbert space has a dimension 4$^N$ 
$=(\sqrt{2}\times\sqrt{2}\times\sqrt{2}\times\sqrt{2})^{N}$. State vectors of 
the physical Hilbert space satisfy
the condition,
\begin{eqnarray}
\label{cons1}
D_i\vert\Psi\rangle_{\rm phys}&=&\vert\Psi\rangle_{\rm phys}\\
\label{cons2}
D_i&\equiv&c_i~c^x_i c_i^y c_i^z
\end{eqnarray} 
The spin operators can then be represented by,
\begin{equation}
\label{sprep}
\sigma^a_i=ic_ic_i^a,~~~~~~~a=x,y,z
\end{equation}
When projected into the physical Hilbert space, the operators defined
above satisfy the algebra of spin $1/2$ operators,
$[\sigma^a_i,\sigma^b_j]= i\epsilon_{abc}\sigma^c \delta_{ij}$.
The Hamiltonian written in terms of the Majorana fermions is,
\begin{eqnarray}
H=-\sum_{a=x,y,z}J_a\sum_{\langle ij\rangle_a}~
ic_i{\hat u}_{\langle ij\rangle_a}c_j, 
\end{eqnarray}
with ${\hat u}_{\langle ij\rangle_a}\equiv ic_i^ac_j^a$.
Kitaev showed that $ [H,{\hat u}_{\langle ij\rangle_a}]=0 $ and 
$u_{\langle ij \rangle a}$ become constants of motion with 
eigen-values ${u}_{\langle ij\rangle_a} = \pm 1 $. The variables 
${u}_{\langle ij\rangle_a}$ are identified with static (Ising) $Z_2$ gauge 
fields on the bonds. Kitaev Hamiltonian (equation 6) has a local $Z_2$ 
gauge invariance in the extended Hilbert space. For practical purposes, the 
local $Z_2$ gauge transformation amounts to 
$ u_{\langle ij \rangle a} \rightarrow\tau_i u_{\langle ij \rangle a} \tau_j$, 
with $\tau_i \pm 1$. Equation (\ref{cons1}) is the Gauss law and the physical 
subspace is the gauge invariant sector.

In the gauge field sector we have gauge invariant $Z_2$ vortex charges 
$\pm 1$ (0 and $\pi$-fluxes), defined as product of 
$u_{\langle ij \rangle a}$ around each elementary hexagonal plaqauette.

Equation 6, with conserved ${\hat u}_{\langle ij\rangle_a}$ is the 
Hamiltonian of free Majorana fermions in the background of frozen $Z_2$ 
vortices or $\pi$-fluxes. 
Since $Z_2$ gauge fields have no dynamics, all eigenstates can be written as 
products of a state in the $2^{\frac{1}{2}N}$ dimensional Fock space of  the 
$c_i$ Majorana fermions and the $(2)^{\frac{3}{2}N}$ dimensional space of 
$Z_2$ link variables. We will refer to the former as {\em matter sector} and 
the latter as {\em gauge field sector}. Gauge copies 
(eigen-states with same energy eigen-values) spanning corresponding 
extended Hilbert space are obtained  by local gauge transformations 
$u_{\langle ij \rangle a }\rightarrow \tau_i u_{\langle ij\rangle a}\tau_j$. 

It turns out that if we attempt to calculate spin-spin correlation functions 
with the use of above free Majorana Hamiltonian and the $Z_2$ fields 
${u}_{\langle ij\rangle_a}$'s, it is difficult to proceed further. 

It is here we have invented a simple but key transformation that 
facilitates exact computation of all spin correlation functions. We call this 
as `bond fermion' formation. In the process we also discover a 
`quantum fractionization' phenomenon in the Kitaev Model, that has an 
unusual validity at all energy scales. 

Hereinafter, we follow the convention that $i$ in the bond 
$\langle ij\rangle_a$, belongs to $A$ and $j$ to $B$ sub-lattice.
We define complex fermions on each link as,
\begin{eqnarray}
\label{chiadef1}
\chi_{\langle ij\rangle_a}&=&\frac{1}{2}\left(c_i^a+ic_j^a\right)\\
\label{chiadef2}
\chi^\dagger_{\langle ij\rangle_a}&=&\frac{1}{2}\left(c_i^a-ic_j^a\right)
\end{eqnarray}
The link variables are related to the number operator of these fermions,
${\hat u}_{\langle ij\rangle_a}\equiv ic_i^ac_j^a = 
2\chi^\dagger_{\langle ij\rangle_a}\chi_{\langle ij\rangle_a}-1$.
All eigenstates can therefore be chosen to have a definite 
$\chi$ fermion occupation number. The Hamiltonian is then block
diagonal, each block corresponding to a distinct set of $\chi$
fermion occupation numbers. Thus all eigenstates in the extended Hilbert 
space take the factorized form,
\begin{eqnarray}
\label{es1}
\vert{\tilde \Psi}\rangle =  \vert {\cal M}_{\cal G};{\cal G }\rangle & \equiv &
\vert {{\cal M}_{\cal G}}\rangle
\vert{\cal G }\rangle\\
\label{es2}
{\rm and}~~ \chi^\dagger_{\langle ij\rangle_a}\chi_{\langle ij\rangle_a}
\vert{\cal G} \rangle
&=&n_{\langle ij\rangle_a}
\vert \cal G \rangle 
\end{eqnarray}
where $n_{\langle ij\rangle_a}=\frac{u_{\langle ij\rangle_a}+1}{2}$ and 
$\vert{\cal M}_{{\cal G}}\rangle$ is a many body eigen-state in the matter 
sector, corresponding to a given $Z_2$ field of $ \vert {\cal G} \rangle $. 
In terms of bond fermions, spin operators become,
\begin{eqnarray}
\label{spchi1}
\sigma_i^a&=&ic_i\left(\chi_{\langle ij\rangle_a}
+\chi^\dagger_{\langle ij\rangle_a}\right)\\
\label{spchi2}
\sigma_j^a&=& c_j\left(\chi_{\langle ij\rangle_a}
-\chi^\dagger_{\langle ij\rangle_a}\right)
\end{eqnarray}
Three components of a spin operator at a site, gets connected to three 
different Majorana fermions defined on the three different bonds !
Written in the above form, the effect of $\sigma^a_i$ on any eigen-state,
which we refer to as a ''spin flip", 
becomes clear. In addition to adding a Majorana fermion at site i, it changes 
the bond fermion number from 0 to 1 and vice versa 
(equivalently, $u_{\langle ij \rangle a} \rightarrow - ~ u_{\langle ij \rangle 
a}$), at the bond $\langle ij \rangle a$. The end result is that one $\pi$ 
flux each is added to two plaquettes that are shared by the bond 
$\langle ij \rangle a$ (figure 2). We denote this symbolically as 
\begin{eqnarray}
\label{pinot}
\sigma_i^a = ic_i\left(\chi_{\langle ij\rangle_a}
+\chi^\dagger_{\langle ij\rangle_a}\right) ~~\rightarrow ~~ ic_i~
{\hat\pi}_{1\langle ij \rangle a}~{\hat\pi}_{2\langle ij \rangle a}~~~~
\end{eqnarray}
with ${\hat \pi}_{1\langle ij \rangle a}$ and 
${\hat \pi}_{2\langle ij \rangle a}$
defined as operators that add $\pi$ fluxes to plaquettes 1 and 2 shared
by a bond $\langle ij \rangle a$ (figure 2). Further 
${\hat \pi}_{1\langle ij \rangle a}^2 = 1 $,
since adding two $\pi$ fluxes is equivalent to adding (modulo $2\pi$) zero flux.

Now we wish to calculate spin-spin correlation functions in physical subspace. 
{\textit {Since the spin operators are gauge invariant, we can compute the correlation
in any gauge fixed sector and the answer will be the same as in the physical
gauge invariant subspace}}. (We have confirmed this by a calculation in the projected 
physical subspace.) So we consider the 2-spin dynamical correlation 
functions, in an arbitrary eigen-state of the Kitaev Hamiltonian in some fixed
gauge field configuration ${\cal G}$,
\begin{equation}
\label{d2scf1}
S_{ij}^{ab}(t)= \langle{\cal M}_{\cal G}\vert\langle{\cal G}\vert 
\sigma_i^a(t)\sigma_j^b(0)
\vert{\cal G}\rangle\vert{\cal M}_{\cal G}\rangle
\end{equation}
Here $A(t) \equiv e^{iHt} A e^{-iHt}$ is the Heisenberg representation of an 
operator A, keeping $\hbar = 1$. As discussed above, 
\begin{eqnarray}
\label{sigones1}
\sigma_j^b(0)\vert{\cal G}\rangle\vert{\cal M}_{\cal G}\rangle&=&
c_i(0)\vert{\cal G}^{ia}\rangle\vert{\cal M}_{\cal G}\rangle\\
\label{sigones2}
\sigma_i^a(t)\vert{\cal G}\rangle\vert{\cal M}_{\cal G}\rangle&=&
e^{i(H-E)t}c_j(0)\vert{\cal G}^{jb}\rangle\vert{\cal M}_{\cal G}\rangle
\end{eqnarray}
where, $\vert{\cal G}^{ia(jb)}\rangle$ denote the states with extra
$\pi$ fluxes added to ${\cal G}$ on the two plaquettes adjoining the 
bond $\langle ik\rangle_{a}\rangle(\langle lj\rangle_{b}\rangle)$ and
$E$ is the energy eigenvalue of the eigenstate 
$\vert{\cal G}\vert{\cal M}_{\cal G}\rangle$. Since the $Z_2$ fluxes on 
each plaquette is a conserved quantity, it is clear that the correlation
function in equation(\ref{d2scf1}) which is the overlap of the two 
states in equations (\ref{sigones1}, \ref{sigones2}) is zero unless 
the spins are on neighbouring sites. Namely, we have proved that the
dynamical spin-spin correlation has the form,
\begin{eqnarray}
\label{d2scfres1}
S_{ij}^{ab}(t)&=&g_{\langle ij \rangle_a}(t)\delta_{a,b}~, ~~~{ij~~ 
\rm nearest ~neighbors}~~ \\ \nonumber
&=& 0 ~~~~~~~~~~~~~~~~~~~{\rm otherwise. } 
\end{eqnarray}
Computation of $g_{ij}(0)$ is straight forward 
in any eigen-state $|\cal M_{\cal G}\rangle$. For the ground state 
where conserved $Z_2$ charges are unity at all plaquettes, the equal time 
2-spin correlation function for the 
bond $\langle ij \rangle a$ is given by the analytic expression:
\begin{eqnarray}
\nonumber
\langle \sigma^{a}_{i}\sigma^{a}_{j}\rangle \equiv
S^{aa}_{\langle ij \rangle a}(0)
=\frac{\sqrt{3}}{16\pi^{2}}\int_{BZ} \cos\theta(k_{1},k_{2})dk_{1}dk_{2}
\end{eqnarray}
Where $\cos\theta(k_{1},k_{2})=\frac{\epsilon_{k}}{E_{k}}$, 
$ E_{k}=\sqrt{(\epsilon_{k}^{2}+\Delta^{2}_{k})}$, in the Brillouin zone. 
$\epsilon_{k}=2(J_{x} \cos k_{1} +J_{y}\cos k_{2} +J_{z})$,  
$\Delta_{k}=2(J_{x}\sin k_{1}+J_{y}\sin k_{2})$,
$k_{1}=\textbf{k}.\textbf{n}_1$, $k_{2}=\textbf{k}.\textbf{n}_2$ and 
$\textbf{n}_{1,2}=\frac{1}{2}\textbf{e}_{x} \pm
\frac{\sqrt{3}}{2}{\textbf{e}_{y}}$ are unit vectors along x and y type bonds.
At the point, $J_x=J_y=J_z$, we get 
$S^{aa}_{\langle ij \rangle a}(0) = -0.52 $.

To compute $g_{\langle ij\rangle_a}(t)$ we substitute for the $\sigma$'s 
from equation (\ref{chiadef1}) and (\ref{chiadef2}). We choose 
a gauge  where $u_{\langle ij\rangle_a}=-1$ implying 
$\chi^{\dagger}_{\langle lj\rangle_{b}}~\vert{\cal G}\rangle=
\chi^{\dagger}_{\langle ik\rangle_{b}}~\vert{\cal G}\rangle=0$.
We note that the above conditions imposed at $t=0$ will continue
to be true at all times since the bond fermion numbers are conserved. 
We then have,
\begin{eqnarray}
g_{\langle ij\rangle a}(t)&=&
\langle{\cal M_{\cal G}\vert}\langle{\cal G}\vert ~ ic_{i}(t)
\chi^{\dagger}_{\langle ij\rangle_{a}}(t)\chi_{\langle ij\rangle_{a}}(0) 
~c_j(0)\vert{\cal G}\rangle \vert{\cal M_{\cal G}}\rangle
\nonumber\\
\end{eqnarray}
The time dependence evolution can be  expressed in terms of the hamiltonian
and noting it is diagonal in the number operators $\chi^\dagger\chi$, we get,
\begin{equation}
\label{time0}
g_{\langle ij\rangle a}(t)=\langle {\cal M}_{\cal G}\vert 
e^{iH[{\cal G}^{ia}]t} i c_i(0)e^{-iH[{\cal G}^{ia}]t}(-1)c_j(0)
\vert {\cal M}_{\cal G}\rangle
\end{equation}
where $H[{\cal G}^{ia}]$ is the tight binding hamiltonian in the background
of the static gauge field configuration ${\cal G}^{ia}$. The $(-1)$ factor
is $u_\langle ij\rangle_a$. This expression can be written in terms of 
the time evolution under $H[{\cal G}]$ as follows, 
\begin{eqnarray}
\nonumber
g_{\langle ij\rangle_a}(t)&=&
\langle{\cal M_{\cal G} \vert}i c_i(t)
T\left(e^{-2J_a \int^{t}_{0}
u_{\langle ij\rangle_a}c_{i}(\tau)c_{j}(\tau)d\tau}
\right)\\
\label{time2}
&&~~~~~~~~~~~u_{\langle ij\rangle_a}c_j(0)
\vert{\cal M_{\cal G}}\rangle
\end{eqnarray}
The above equation is written in an arbitrary gauge.

We have thus derived a simple but exact expression for the spatial dependence 
of the two spin dynamical correlation function.  We have also obtained an 
exact expression for the time dependence in terms of the correlation
functions of non-interacting Majorana fermions in the background of 
static $Z_2$ gauge fields. Equation  (\ref{time2}) represents the propagation
of a Majorana fermion in the presence of two injected fluxes. It can be treated
as an X-ray edge problem and computed in terms of the Toeplitz determinant.
We will not do this now but proceed to discuss some general features of 
our results.

\begin{figure}[h!]
\center{\hbox{\epsfig{figure=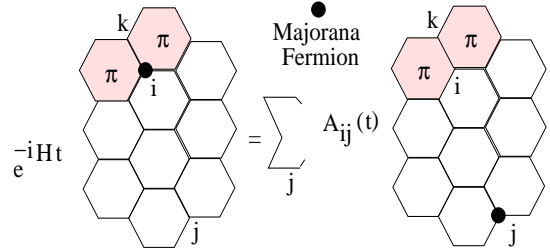,height=1.3in,width=2.8in}}}
\caption{Time evolution and fractionization of a spin flip at t = 0 at 
site i, into a  $\pi$-flux pair and a propagating Majorana fermion. 
`Shakeup' of the Majorana fermion vacuum to an instantaneous addition 
at t = 0, of a $\pi$-flux pair is not shown.}
\end{figure}

The notion of fractionalization of spin-flip quanta is the natural
interpretation of our results \cite{frac,bza}. Consider time evolution of a single 
`spin-flip' at site i given in equation (\ref{sigones2}).
Using the notation introduced in equation (\ref{pinot}) we have,
\begin{eqnarray}
\sigma_i^a \vert {\hat \Psi} \rangle \equiv
i c^{}_{i}(t) T(e^{2u_{\langle ik \rangle a}J_{a}\int^{t}_{0}c^{}_{i}(\tau)c^{}_{k}(\tau)d\tau})
{\hat \pi}_{\langle ik \rangle a 1} {\hat \pi}_{\langle ik \rangle a 2}~\vert {\hat \Psi} \rangle
\nonumber\\
\end{eqnarray}
A spin-flip at site i at time t = 0 is a sudden perturbation to the matter (Majorana fermion)
sector, as it adds two static $\pi$-fluxes to adjoining plaquettes. The time ordered
expression represents how a bond perturbation term, $i2u_{\langle ik \rangle a}J_{a}c^{}_{i}c^{}_{k}$ evolves the Majorana fermion state, in `interaction representation'. At long time scale the resulting `shakeup' is simple and represents a rearrangement (power law type for gapless case) of the Majorana fermion vacuum to added static $\pi$-flux pairs. The Majorana fermion, produced by a spin-flip, $c_i(t)$ propagates freely, as a function of time. 

As spin-flip at site i is a composite of a Majorana fermion and 
$\pi$-flux pair (equation 13), two spin correlation function defines the 
probability that we will detect the added composite at site j after a time t. 
As the added $\pi$-flux pair do not move, the above probability is identically 
zero, unless sites i and j are nearest neighbors and spin components are a = b. 
This is why the spatial dependence of two spin correlation functions are 
sharply cut off at nearest neighbor separation. The asymptotic response to an 
added $\pi$-flux pair and free dynamics of the added Majorana fermion control 
the long time power law behaviour of our only non vanishing nearest neighbor 
two spin correlation function.

Further, for a given pair of nearest neighbor sites, only one Ising spin pair
of a corresponding component is non-zero. Other pairs and cross correlation 
functions vanish. More specifically, for a given bond the only non zero 
two spin correlation function is the bond energy. 

What is unusual is that the above result is true in all eigen-states of the 
Kitaev Model, irrespective of energies. It follows that it is valid for 
thermal averages too. This is an unusual result, indicating exact 
fractionalization occurring at all energy scales. In known models such as 
1D repulsive Hubbard model or spin half Heisenberg chain, fractionization is 
only a low energy asymptotic phenomenon. Our results show the 
\emph{all energy scale exact confinement} of the spin-flip quanta, and exact 
\emph{deconfinement} of the Majorana fermions in the Kitaev model.

It is interesting to see that the above is a special property of the 
Kitaev Model. When we perturb it by adding, for example, a magnetic field term 
or make bond terms non-Ising, $\pi$-fluxes acquire dynamics. This means that 
the probability amplitude of finding the composite particle intact at a 
farther site is finite (though exponentially small as a function of separation) 
and not strictly zero.

Multi spin correlation functions can be calculated in our formalism. Further, 
quantum entanglement, a key notion in quantum computation and quantum 
information, is ultimately connected with some complicated multi-spin 
correlation function. We have computed some entanglement measures, 
but do not discuss them in the present paper.

To summarise, this paper presents certain exact analytical results for the 
spin dynamics and a spin-flip fractionization scheme for the Kitaev Model. 
As this non-trivial spin model is also a model for topological quantum 
computation, our exact results should provide insights into qubit dynamics 
and possible ways of generating emergent topological qubits. Our formalism, 
which uses the factorized character of the eigen-functions in the extended 
Hilbert space, is easily adapted to the calculation of multi-spin correlation 
functions, which is a key step in the calculation and understanding of 
quantum entanglement properties. 

\begin{flushleft}
\bf {Acknowledgement}
\end{flushleft}
G.B. thanks Ashvin Vishwanath for bringing the Kitaev Model to his attention and
tutorials.

\vfill
\end{document}